\documentclass[reprint, amsmath,amssymb, aps, superscriptaddress, floatfix]{revtex4-2}

\usepackage[margin=1in]{geometry}
\usepackage{amsmath, amssymb}
\usepackage{graphicx}
\usepackage[normalem]{ulem}

\let\Im\undefined
\DeclareMathOperator{\Im}{Im}
\DeclareMathOperator{\Tr}{Tr}

\usepackage{xcolor}

\begin{document}

\title{Gating-Induced Mott Transition in NiS$_2$}

\author{Ezra Day-Roberts}
\affiliation{School of Physics and Astronomy, University of Minnesota, Minneapolis,
MN 55455, USA}
\affiliation{Department of Chemical Engineering and Materials Science, University of Minnesota, MN 55455, USA}
\author{Rafael M. Fernandes}
\affiliation{School of Physics and Astronomy, University of Minnesota, Minneapolis,
MN 55455, USA}
\author{Turan Birol}
\affiliation{Department of Chemical Engineering and Materials Science, University of Minnesota, MN 55455, USA}
\date{\today}

\begin{abstract}
    NiS$_2$ has been widely regarded as a model system to study the bandwidth-controlled Mott transition, as enabled by isovalent Se chemical substitution on the S sites. Motivated by advances in electrostatic gating, we theoretically investigate the filling-controlled Mott transition induced by gating, which has the advantage of avoiding disorder introduced by dopants and stoichiometric changes. We use combined Density Functional Theory (DFT) and Dynamical Mean Field Theory (DMFT) to study such a filling-controlled transition and compare it with the case of bandwidth control. We draw a temperature-filling phase diagram and find that the Mott-insulator to metal transition occurs with modest added electron concentrations, well within the capabilities of existing electrostatic gating experiments. We find that there is significant incoherent weight at the Fermi level in the metallic phase when the transition is induced by gating. In contrast, the spectral weight remains rather coherent in the case of the bandwidth-controlled transition.
\end{abstract}

\maketitle

\section{Introduction}
While a simple quantum mechanical description of non-interacting electrons is sufficient to explain the insulating phase of materials with fully occupied atomic orbitals \cite{Slater1934}, it was realized early on that interactions between electrons need to be taken into account to explain the insulating nature of compounds with partially filled transition metal d orbitals such as NiO \cite{Mott1937, Ashcroft1976}. In these so-called Mott insulators, the onsite screened Coulomb repulsion outcompetes the kinetic energy gained by delocalized electrons, resulting in an insulating phase with localized electrons. 
While most Mott insulators undergo a temperature-driven metal-to-insulator transition (MIT) before they melt \cite{Limelette2003,Kanoda2011,Kotliar2000,Papanikolaou2008,Tremblay2012}, another interesting scenario is that in which the MIT can be triggered at fixed temperature by tuning the material properties \cite{Imada1998}. These transitions are the subject of ongoing interest due to both a plethora of interesting emergent phenomena associated with the MIT, including possible quantum critical behavior \cite{Dobrosavljevic2012,Vlad2013,Vojta2019,Mak2021}. 

At a fixed temperature, metal-insulator transitions can be induced by changes in the electron-electron interaction strength, the bandwidth, or the band filling, i.e. the number of valence electrons per atom \cite{Dobrosavljevic2012}. While theoretical models enable studying the effects of all of these parameters separately, experiments typically focus on the latter two \cite{Imada1998}. 
Bandwidth is typically tuned by either hydrostatic or chemical pressure, which refers to the isovalent substitution of atoms that preferably do not have any electronic states near the Fermi level. A typical example of chemical pressure is substitution of the A-site cations in perovskite oxides, which are typically either alkaline earth or rare earth elements, with other isovalent alkaline earth or rare earth atoms. (For example, see \cite{Greedan1985}.) These substitutions changes the unit cell volume and internal structural parameters, but do not affect the number of electrons in the bands near the Fermi level. Anion substitution, such as sulfurization of oxides, can also be used to induce chemical pressure and change the crystal structural parameters. (For example, see \cite{Benjamin2021, Han2016}.) But a more important effect of anion substitution in transition metal oxides is the change in the hopping amplitude $t$ between transition metal orbitals, which modifies the bandwidth near the Fermi level.

Aliovalent doping changes the number of electrons, and is the most commonly used means to change the electron filling. For example, substituting the rare earth atoms in rare earth titanate perovskites with alkaline earth elements adds a hole, which can move the system away from integer-filling, and hence lead to an metallic phase \cite{Katsufuji1997}. However, this approach has two shortcomings when used to study filling-induced Mott transitions: 1) Since no two ions have exactly the same radius, chemical doping necessarily has a steric effect that changes the crystal structure. Hence, it is not solely a change of electron filling. 2) Partial substitution of an element with another rarely results in an ordered crystal structure. Instead, a random  distribution of dopant atoms is more common. This results in local disorder in both the crystal structure and the electrostatic potential the electrons experience \cite{Aguiar2005,Miranda2020}. 

Recently, electrolyte gating approaches that rely on ionic liquids or gels have emerged as an alternate means to modify the carrier concentration and induce phase transitions in solids \cite{Goldman2014, Bisri2017, Leighton2019}. By using an electric double-layer transistor device, it is possible to obtain electrostatic as well as electrochemical effects and induce charge densities that are well beyond what can be achieved by using solid dielectrics \cite{Yuan2009, Leighton2022}. This approach was shown to be successful in, for example, modifying the superconducting properties of SrTiO$_3$ \cite{Lee2011}, KTaO$_3$ \cite{Ueno2011}, and various cuprates \cite{Bollinger2011, Leng2012, Garcia-Barriocanal2013}; inducing metal-insulator transitions in WO$_3$ \cite{Leng2017} and nickelate perovskites \cite{Bubel2015,Scherwitzl2010}; or inducing ferromagnetism in diamagnetic FeS$_2$ \cite{day2020contrasting, Walter_Voltage}. In addition to changing the band filling without necessarily introducing chemical or structural inhomogeneities, gating approaches also help bridge the miscibility gaps in compounds where suitable dopants are not easily identified. 

\begin{figure}
\includegraphics[width=0.8\linewidth]{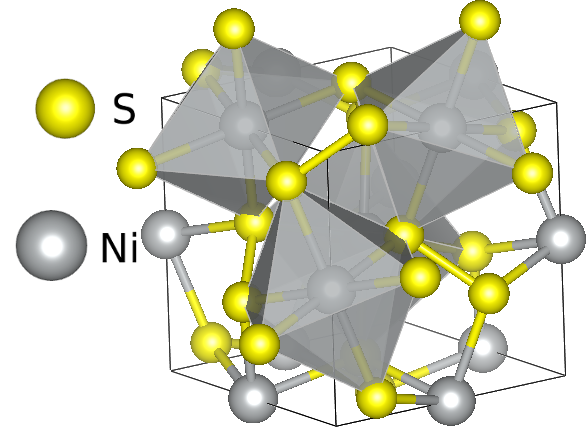}
\caption{The pyrite crystal structure. Grey and yellow atoms represent the transition metal (in this case Ni) and sulfur ions respectively. The Ni--S and S--S bonds have lengths of $2.39\text{ \AA}$ and $2.07\text{ \AA}$ respectively.}
\label{fig:structure}
\end{figure}

NiS$_2$ is a well studied Mott insulator, which provides a fertile playground to examine the metal-insulator transition. It has the pyrite structure named after FeS$_2$ (Fig.~\ref{fig:structure}) \cite{Stillwell1936, Bither1966}. With its cubic symmetry and simple stoichiometry, as well as the availability of large single crystals and solid solutions \cite{Bouchard1968b, Bouchard1968a}, it was considered as a model system to study the Mott transition as early as the 1970s \cite{Jarrett1973, Ogawa1974}.
Sulfur is a chalcogen one row below oxygen in the periodic table and similar to oxygen the most common anion of sulfur is S$^{2-}$ sulfide; however, in the pyrite structure, sulfur ions form S-S dimers, which have a total charge of $-2$ per dimer. The S-S distance is 2.07 \AA, closer to elemental $S_2$ with $d_{S-S}=1.887$\AA\cite{Brownridge2005} and far from the $S-S$ distances in compounds without S-S bonding, such as the perovskite BaZrS$_3$ with $d_{SS}=3.58$\AA. One unusual feature of this structure is the existence of covalently bonded chalcogen dimers, which are marked with bonds in Fig.~\ref{fig:structure}. In contrast to most oxide structures where the chalcogens each have a charge of $-2$, here the sulfur dimer collectively has a charge of $-2$, giving each sulfur an individual ionization of $-1$. In addition to the filling-controlled transition achieved by doping the transition metal site, a bandwidth-controlled transition can be achieved by partially substituting S by Se \cite{Gautier1972, Bouchard1973}. Note that NiSe$_2$ is a Pauli paramagnetic metal, whereas NiS$_2$ is a Curie-Weiss paramagnet that orders antiferromagnetically below 38~K \cite{Bither1968, Wang1997x, Yano2016}. There is now a large body of work focused on NiS$_{2-x}$Se$_x$ that  includes both experimental and theoretical studies \cite{Folkerts1987, Fujimori1996, Iwaya2004,Yao1996,Kunes2010,Han2018x,Moon2015, Wilson1985}. The body of first-principles work on NiS$_{2-x}$Se$_x$ includes a number of Dynamical Mean Field Theory (DMFT) studies, starting from as early as 1998 \cite{Matsuura1998}, and most recently, the discussion of Hund metal behavior near the transition \cite{Jang2021}. 

In this paper, we study the metal-insulator transition in electron-doped NiS$_2$ using first principles density functional theory plus embedded dynamical mean field theory (DFT+eDMFT) \cite{Kotliar2006, Haule2010, Paul2019Rev}. In particular, we consider the electrostatically or electrolyte gated system without chemical doping or electrochamical changes, where the electron concentration increases without any change in the stoichiometry of the system. We thus focus on the carrier filling induced Mott transition, which we compare with the bandwidth-induced transition in the Se substituted system. The behavior of this system under electrostatic gating is of interest  because of the successful application of the gating approach to the closely related compound FeS$_2$ \cite{Walter_Voltage}, as well as the recent experiments on NiS$_2$ that demonstrate a metal-insulator transition under gating \cite{Hameed2022}. However, the role of electrostatic-only effects in these experiments is unclear \cite{Leighton2022}, and a theoretical answer to the question of whether such a transition is feasible in the experimentally achievable electron concentrations is missing so far.  

Our results show that a very small electron concentration, well within the reach of electrolyte gating, is sufficient to suppress the Mott insulating phase. By computing the spectral density and the quasi-particle spectral weight, we also find that the filling induced transition is different from the one induced by changing the bandwidth in key points. Specifically, the former leads to a highly incoherent metallic phase and to features in the DOS reminiscent of the Mott gap. This is in contrast to the bandwidth induced transition, where the gap completely collapses in the metallic phase. 

The paper is organized as follows: Sec. II introduces the DFT+DMFT methodology. The crystal and electronic structures of undoped NiS$_2$ are discussed in Sec. III, whereas Sec. IV presents the DFT+eDMFT results for the DOS and the quasi-particle spectral weight. Sec.V is devoted to the conclusions.

\section{Methods}
Density Functional Theory + Embedded Dynamical Mean Field Theory (DFT+eDMFT) calculations were performed using the Rutgers eDMFT code \cite{Haule2010,Haule2018}. The density functional theory calculations using the local density approximation were performed using the linearized augmented plane wave method as implemented in Wien2K \cite{Blaha2020}. A $10\times 10\times 10$ k-point grid was used for the 12 atom primitive unit cell of NiS$_2$. The DMFT problem was solved via the continuous time Quantum Monte Carlo solver included in the eDMFT code. The self-energy was directly sampled and its high frequency tail was smoothened for stability as described in  \cite{Hiroshi2017}. The Ni 3d orbitals were treated in DMFT with localized states constructed as described in \cite{Haule2010}. In this unit cell, the local axes of each Ni ion are oriented differently, but all atoms are symmetry equivalent, i.e. there is only one Wyckoff position for Ni, and another one for S. As a result, a single Ni ion was considered as the impurity in the DFT+eDMFT calculations. The Green's function and self energies for the orbitals on the other Ni ions were generated by applying space group operations that interchange the Ni ions. 
All 5 3d orbitals of the Ni ions were included in the impurity problem, even though the $t_{2g}$ manifold is completely filled. 
The on-site Hubbard U and Hund's J parameters were set to 10.7 and 0.7 eV, which were shown in multiple references (for example references \cite{Haule2014, Mandal2019} to be suitable for 3d transition metals for the implementation that we use. This number is larger than the typically used values in DFT+U, as well as other DFT+DMFT implementations, due to the facts that DMFT explicitly takes into account certain screening processes and that our DMFT implementation does not used less localized Wannier orbitals as the correlated orbitals. Hybridization with all the states in a wide energy window that spans $\mp10$~eV around the Fermi level was taken into account, and all transition metal orbitals were explicitly included in the impurity calculation. This latter point also has an effect on the value of U, since taking into account the Feynman diagrams involving the occupied d orbitals explicitly in the impurity calculation requires not reducing the value of U due to screening by those orbitals. 
The `exact' double counting approach was used in all calculations \cite{Haule_Exact_2015}. 

Unless otherwise stated, the experimental crystal structure of NiS$_2$ from Ref. \cite{Nowack_Charge_1991} was used in all calculations. While the added electrons can change bonding and effective ionic radii, and hence change the crystal structure \cite{day2020contrasting, Li2021}, this effect is not expected to be significant at the small concentrations considered at this work, and hence is ignored. The lattice constant for Ni(S$_{0.75}$Se$_{0.25}$)$_2$ was linearly interpolated between endpoints NiS$_2$ and NiSe$_2$ and the atomic positions were determined by relaxation in Wien2K (at the DFT level).

The fully self-consistent DFT+DMFT implementation we use does not allow changing the number of electrons independently in the DFT and the DMFT loops. We added electrons to the system by changing the total number of electrons in the DFT calculation along with a homogeneous background charge to keep the system charge neutral \cite{Blaha2001}. The added electrons lead to a change in the impurity occupation in the DMFT calculation.

\section{Crystal and Electronic Structures of {N\lowercase{i}S$_2$}: DFT results}

The pyrite crystal structure shown in Fig.~\ref{fig:structure} is commonly found in the 3d transition metal disulfides, other examples being FeS$_2$, CuS$_2$, and CoS$_2$ \cite{Ogawa1976}. 
Despite its primitive cell that contains 4 formula units, the pyrite structure is highly symmetric and has the simple cubic space group Pa$\overline{3}$ (\#205) \cite{Nowack_Charge_1991}. The transition metal ions occupy the corners and the face-centers of the cubic unit cell, or equivalently the edge-centers if a shifted origin is chosen. Sulfur ions form octahedra around each transition metal. The orientations of these octahedra are not the same with respect to each other, which leads to the simple cubic symmetry rather than the face-centered cubic one. The structure is centrosymmetric with various screw axes and glide planes, and there are 3-fold roto-inversion axes along all body-diagonals. Due to these symmetry operations, every transition metal ion is chemically equivalent, with different local axes, i.e., they occupy a single Wyckoff position 4a. Similarly, all sulfur ions are chemically equivalent.

The previously mentioned covalent bonding in the sulfur dimers gives rise to unoccupied antibonding sulfur bands near the Fermi level \cite{Bither1968, Goodenough1971}. The octahedral coordination environment of the transition metal cations gives rise to a significant $t_{2g}$-$e_g$ splitting of their d orbitals. This splitting is responsible, for example, for the band-insulating behavior of FeS$_2$, as Fe has completely filled $t_{2g}$ and completely empty $e_g$ orbitals \cite{Zhao1993}. 
The Ni$^{2+}$ ions in NiS$_2$ have 8 electrons in their partially filled d-shell. This results in completely filled $t_{2g}$ orbitals and half-filled $e_g$ orbitals \cite{Goodenough1971}. 

\begin{figure}
\includegraphics[width=0.80\linewidth]{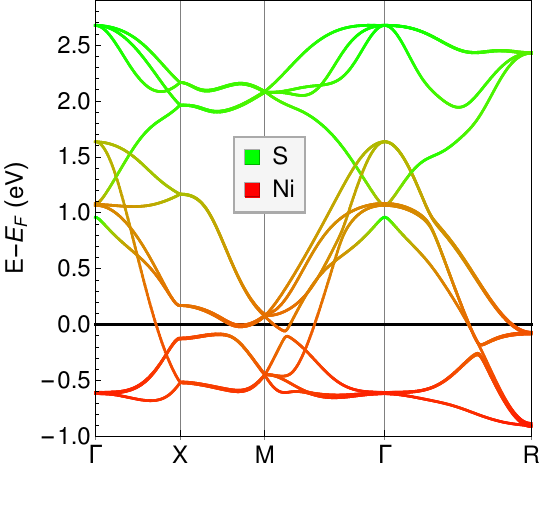}
\caption{Band structure of nonmagnetic NiS$_2$ from density functional theory. The wavefuction of each band at each wavevector is projected onto Ni and S states to obtain the atomic characters of the bands. The higher lying bands arise from the sulfur dimer anti-bonding orbitals, and hence are almost purely of S character. However, these bands overlap with the bands with Ni-e$_g$ character, mixing with them at lower energies. }
\label{fig:DFT}
\end{figure}

The non-spin polarized DFT band-structure of NiS$_2$ is shown in Fig.~\ref{fig:DFT}. Without magnetic moments or a +U correction, DFT predicts this material to be metallic. The Fermi level crosses the 8-band  Ni-$e_g$ manifold, which is separated in energy from the Ni-$t_{2g}$ and occupied sulfur $p$ bands. The sulfur anti-bonding orbitals give rise to 4 unfilled bands, which are located in the range 1.0-2.5~eV above the Fermi level. Because these bands overlap and mix with the Ni-$e_g$ bands, even the lower energy Ni-$e_g$ bands have significant sulfur character, indicating the relatively covalent nature of this material. 


At $T_N\sim 38$~K, NiS$_2$ undergoes a phase transition to a noncollinear antiferromagnetic insulating phase with $S=1$ moments formed by the parallel alignment of the spins of the two electrons on the Ni $e_g$ orbitals \cite{Hastings1970, Higo2015, Yano2016}. While it remains an insulator in the paramagnetic phase, a Mott-insulator to metal transition can be induced by doping \cite{Ogawa1974}, pressure \cite{Wilson1971, Friedemann2016}, or isovalent substitution of S with Se \cite{Gautier1972, Jarrett1973, Wilson1985}. The metal-insulator transition induced by Se substitution has been studied in great detail both theoretically and experimentally \cite{Folkerts1987, Fujimori1996, Iwaya2004,Yao1996,Kunes2010,Han2018x,Moon2015, Wilson1985, Imada1998,Matsuura1998}, and is considered to be a relatively clean example of bandwidth-controlled Mott transition. This is because Se is a neighbor of S in the periodic table, such that its substitution is not expected to introduce strong disorder on the transition metal site.

\section{Correlated electronic spectrum: DFT+$\mathrm{e}$DMFT Results}

\begin{figure}[htp]
\includegraphics[width=0.9\linewidth]{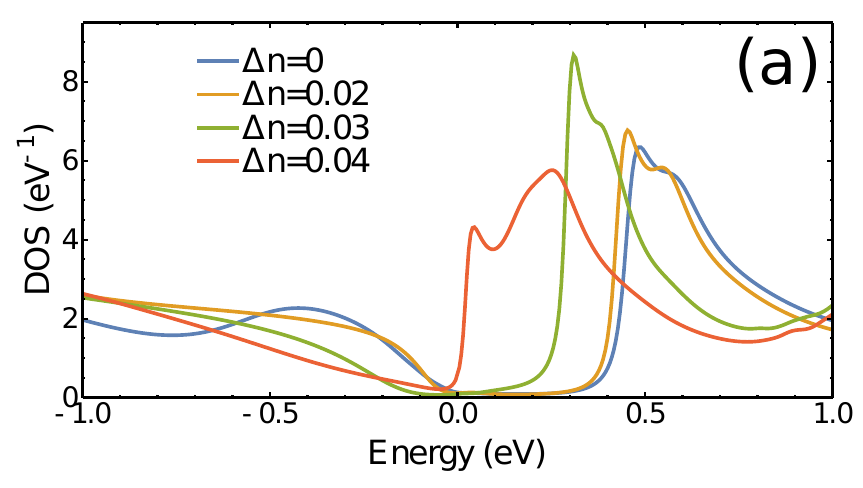}
\includegraphics[width=0.9\linewidth]{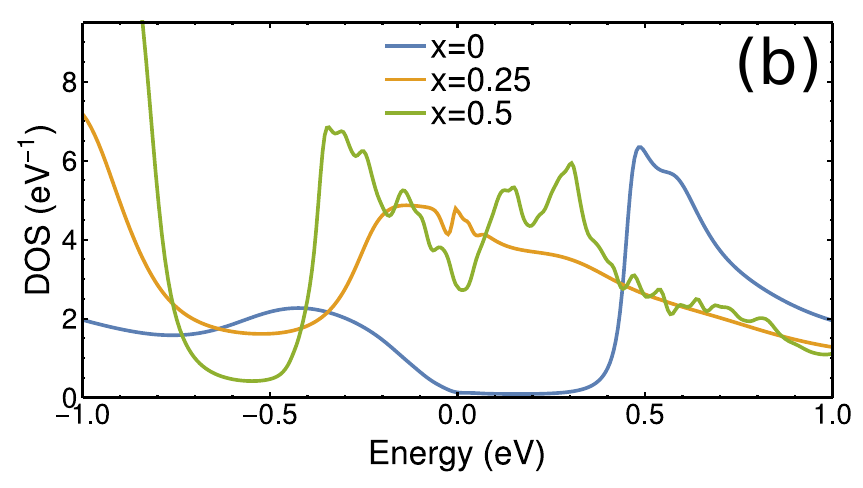}
\caption{Evolution of the DOS of NiS$_2$ under electrostatic gating (a) and Se substitution (b). Panel (a), obtained at a constant temperature of $T=223$~K, shows that the Mott gap slowly closes upon increasing the number of electrons per Ni ion, $\Delta n$. At $\Delta n=0.04$ the Fermi level has moved into the upper Hubbard band, but there is still a clear suppression of the DOS immediately below it. Panel (b) refers to the isovalent substituted Ni$($S$_{1-x}$Se$_x)_2$ at the same temperature. Note that the Mott  gap completely collapses for $x=0.25$.}
\label{fig:DOS}
\end{figure}

In Fig.~\ref{fig:DOS}a, we show the density of states (DOS) of NiS$_2$ obtained from DFT+eDMFT calculations as a function of electron doping induced by electrostatic gating. Doping is simulated by adding electrons at the DFT level and allowing this to propagate to the DMFT portion via full charge self-consistency. The number of electrons in the unit cell is modified in Wien2k along with the addition of a homogeneous background charge to maintain charge neutrality. At the temperature we consider ($T=223$~K), undoped NiS$_2$ is a Mott insulator with a well defined gap and minimal incoherent weight at this gap. Increasing the number of electrons as much as $\Delta n=0.03$~e$^-$ per Ni ion shifts the Fermi level up while at the same time increasing the DOS in the gap slightly, even though a gap remains visible. The Fermi level $E_F$ crosses the bottom of the conduction band and the DOS at $E_F$ becomes significant when $\Delta n=0.04$, suggesting the onset of metallic behavior. Nevertheless, a dip in the DOS below the Fermi level persists, resembling a remnant gap. These features are resilient to details of analytic continuation, see the supplemental materials, figure S.1. This behavior is in contrast to the bandwidth-controlled MIT in Ni$($S$_{1-x}$Se$_x)_2$. As shown in figure \ref{fig:DOS}b, even when 25\% of S atoms are substituted with isovalent Se, the Mott gap completely collapses, and any feature resembling a gap is completely absent. The latter is consistent with what was previously reported in Ref.~\cite{Moon2015}. 

\begin{figure*}[htp]
\centering
\includegraphics[width=0.9\textwidth]{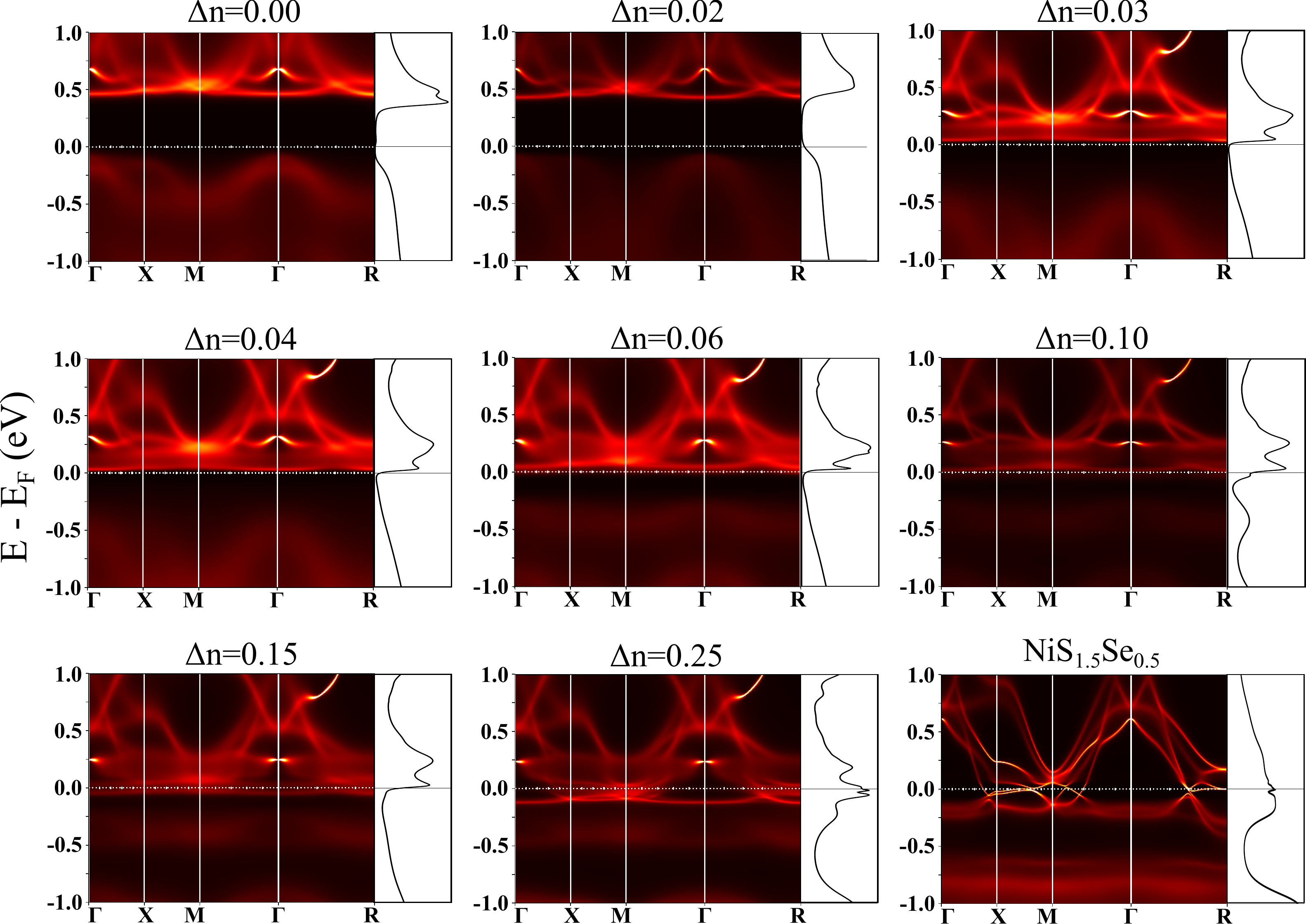}
 \caption[Various spectral functions]{Evolution of the spectral function $A(k,\omega)$ of NiS$_2$ as a function of increasing electron concentration $\Delta n$, and also for the case of 25\% Se substitution. All plots are from calculations performed at $T=223$~K. The DOS curves are shown in the right side of each panel. According to the self-energies shown in Fig.~\ref{fig:sigs}, the systems shown in the first three plots ($\Delta n\leq0.03$) exhibit insulating behavior.}
\label{fig:Akw}
\end{figure*}

To gain further insight into the differences between filling- (gating) and bandwidth- (Se substitution) controlled Mott transitions, we show in Fig.~\ref{fig:Akw} the spectral functions $A(k,\omega)$ in both situations. In the case of doping, the Hubbard band at the bottom of the conduction manifold remains nearly unchanged with increasing $\Delta n$, despite the fact that the gap in the DOS is slowly filled with incoherent states. 
Even for $\Delta n>0.06$, where the the DOS is non-zero at the Fermi level, and the Fermi level starts crossing the upper Hubbard band of the undoped compound, there are no clearly defined dispersive bands. This is consistent with a correlated metallic state where most of the spectral weight is incoherent, including near the Fermi level. In contrast, the 25\% Se substituted system has sharp bands crossing the Fermi level, indicative of a coherent metallic regime.

To more quantitatively assess these behaviors, we show the imaginary part of the self-energy on the Matsubara axis, $\Im \Sigma(i\omega)$, as it can be used to gauge the strength of electronic correlations in the metallic state. Specifically, following the method of Ref. \cite{Dang2015}, the slope of $\Im \Sigma(i\omega)$ as $\omega \rightarrow 0$ is related to the quasi-particle weight $Z$ via 
\begin{equation}
    Z^{-1}=1 - \left. \frac{\partial \Im \Sigma(i \omega)}{\partial \omega}\right|_{\omega \rightarrow 0} .
\end{equation}
We fit a fourth-degree polynomial to the lowest 5 Matsubara frequencies and extract the linear component. Thus, a large negative slope indicates strong correlations (i.e. a small $Z$), whereas a divergent positive slope signals a Mott insulating phase (i.e. $Z=0$). The advantage of using this expression for calculating $Z$, rather than a similar expression on the real frequency axis, is that it does not require the use of numerical analytical continuation methods such as the maximum entropy method, which are bound to introduce errors. The results from this do not depend sensitively on the degree of polynomial or number of points fit, see supplement figure S.2. 

\begin{figure}[htp]
\includegraphics[width=0.8\linewidth]{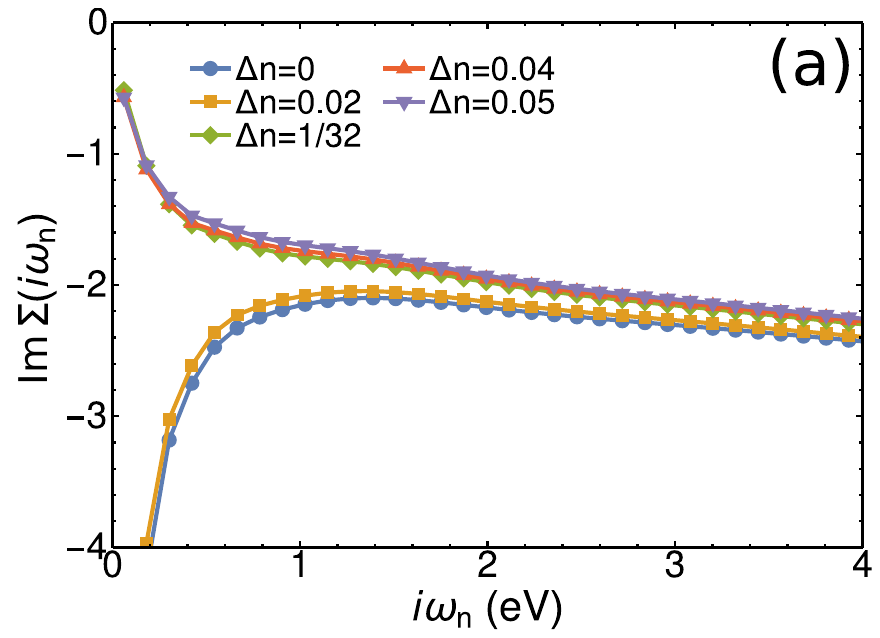}
\includegraphics[width=0.8\linewidth]{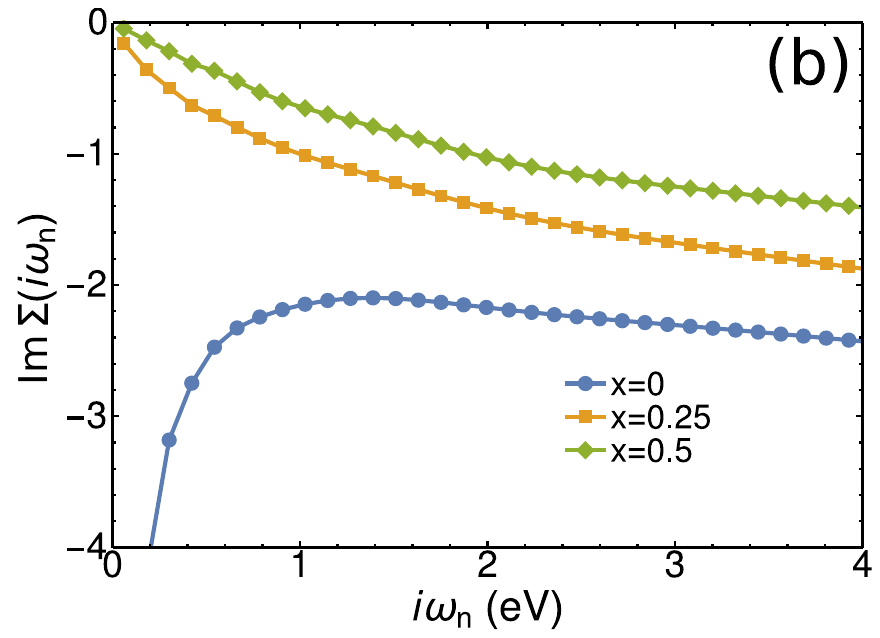}
\caption[Imaginary part of the self-energy of NiS$_2$.]{(a) The imaginary part of the self-energy on the Matsubara axis, $\Im \Sigma(i\omega)$, shows a clear MIT upon increasing electron concentration around the value of $\Delta n=1/32$ added electrons per Ni ion. Even in the metallic state, the compound shows strong electron-electron scattering and correlations as evidenced by the non-zero intercept of $\Im \Sigma(i\omega)$ and its large slope approaching zero Matsubara frequency. (b) The imaginary part of the self-energy of Se-substituted Ni$($S$_{1-x}$Se$_x)_2$ also shows a clear MIT when $x=25\%$. In the metallic state, although the slope of $\Im \Sigma(i\omega)$ indicates the presence of bandwidth renormalization, the compound is a good metal since $\Im \Sigma(i\omega)$ has essentially a zero intercept with the $\omega_n =0$ axis.}
\label{fig:sigs}
\end{figure}

The self energy $\Sigma(i\omega)$ of the $e_g$ orbitals is shown in Fig.~\ref{fig:sigs}a for different values of electron doping $\Delta n$. For smaller $\Delta n$ values, the self-energy is divergent, signalling a Mott insulating phase. For larger doping values, however, $\Sigma(i\omega)$ displays a sizable but doping-independent slope, signalling a correlated metallic phase. The imaginary part of the self-energy in the metallic phase has a non-zero intercept for all metallic doping levels, consistent with large electron-electron scattering and a large degree of remaining incoherent weight. The finite-temperature Mott transition is described by a first-order transition line terminating at a critical endpoint \cite{Kotliar2000,Bulla2001}. Although we are not able to determine this endpoint from our calculations, it is unlikely to extend to high temperatures. Thus, since we see a sharp change in the asymptotic behavior of the self-energy as a function of $\Delta n$, we use this criterion to delineate the metallic and insulating phases in the doping-temperature phase diagram. In comparison to the electron-doping case, the self-energy in the Se substitution case displays a very small intercept (Fig.~\ref{fig:sigs}b), consistent with there being less scattering and less incoherent weight at the Fermi level. Indeed, while the bandwidth is renormalized significantly ($Z=0.22$ for Ni(S$_{0.75}$Se$_{0.25}$)$_2$), the bands in Fig.~\ref{fig:Akw} appear ``sharp" and hence coherent.

\begin{figure}[htp]
\includegraphics[width=0.95\linewidth]{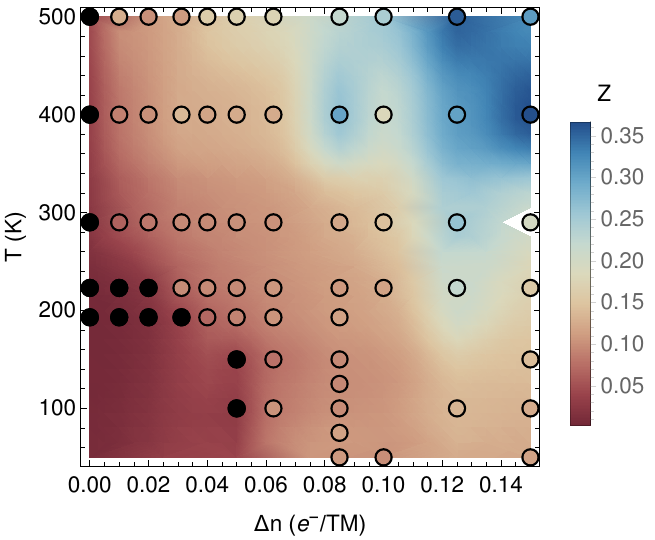}
\includegraphics[width=1.05\linewidth]{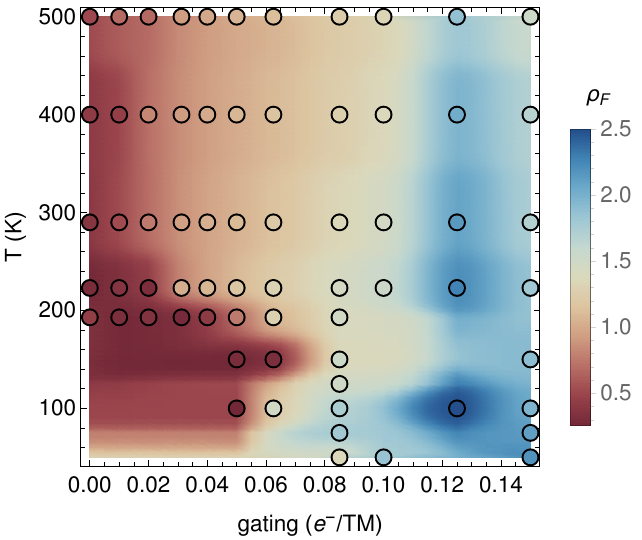}
\caption[Gating-temperature phase diagrams of NiS$_2$ MIT]{Gating (in terms of the number of added electrons per Ni atom)-temperature phase diagrams showing both the quasi-particle spectral weight $Z$ (top), and the proxy to the DOS at the Fermi level $\rho_F$ (bottom), as defined in the main text. Circles indicate the values where the DFT+eDMFT calculations were performed and insulating points in the upper figure are marked with solid black dots; the phase diagram was then obtained using a linear interpolation scheme.}
\label{fig:phaseDiagram}
\end{figure}

In Fig. \ref{fig:phaseDiagram} we present the gating (in terms of the number of added electrons per Ni atom, $\Delta n$) \emph{versus} temperature phase diagram of NiS$_2$. The quasiparticle residue $Z$ is shown together with the imaginary axis Greens's function evaluated at the lowest Matsubara frequency, $\rho_F=-\Im\Tr G(i\omega_0)$, which is a proxy for the density of states at the Fermi level. These two quantities show slightly different trends, but are in good agreement within the error bars. Indeed, the points displaying Mott insulating behavior ($Z=0$) coincide with those with almost zero DOS at the Fermi level, $\rho_F\approx 0$. Undoped NiS$_2$ is insulating at all temperatures up to $500$ K in our calculations, but the temperature range of the insulating region on the phase diagram narrows with the introduction of electrons via gating. The system develops a nonzero $Z$ (and hence becomes  metallic) with the addition of even $0.01$ electrons per f.u. above a temperature of about $\sim\!\!220$~K. Below this temperature, there is a wider insulating region, extending to $\Delta n\approx 0.05$. Above concentrations of $\Delta n\approx0.06$, there is sizable DOS at the Fermi level at all temperatures, whereas $Z$ increases significantly with increasing temperature. Throughout the region spanned by the temperature-doping values we considered, NiS$_2$ remains significantly correlated, with $Z$ remaining below $0.35$.

\begin{figure*}[htp]
\includegraphics[width=0.5\textwidth]{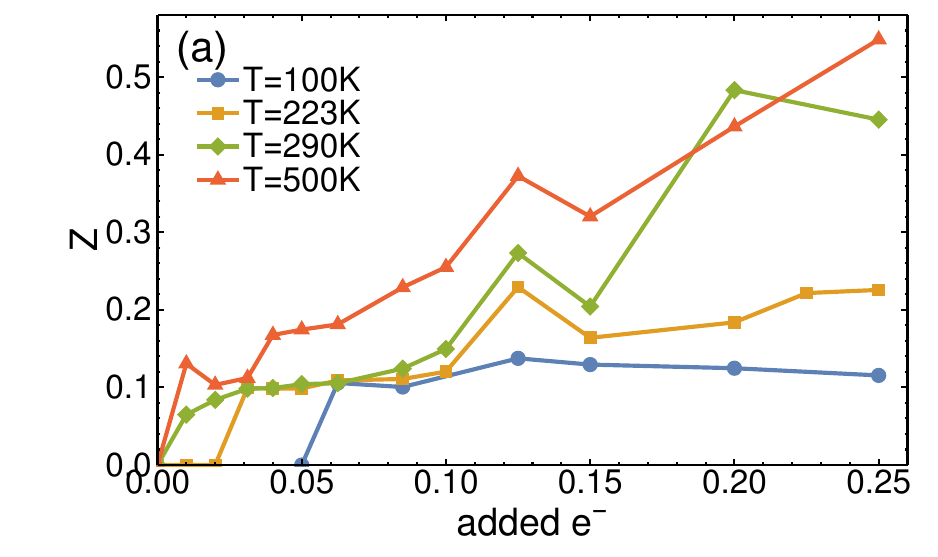}%
\includegraphics[width=0.5\textwidth]{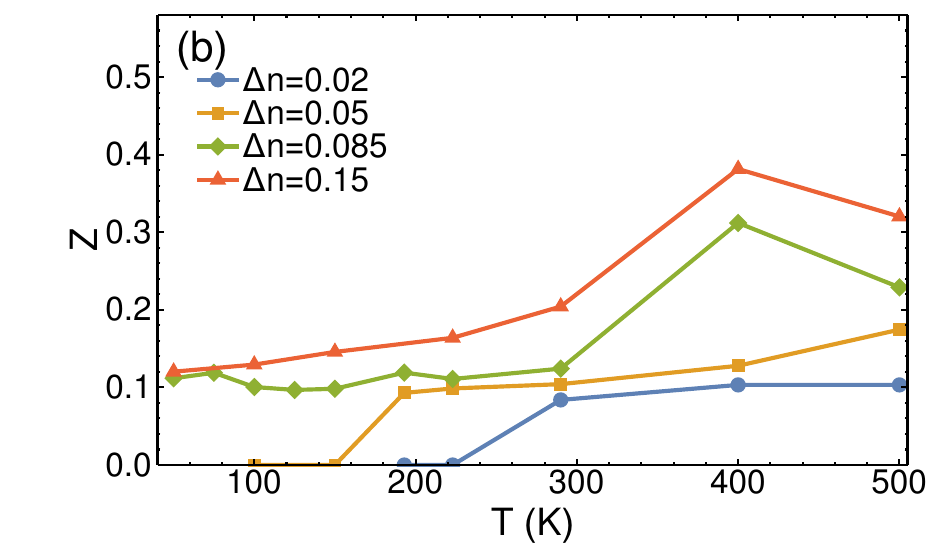}
\includegraphics[width=0.5\textwidth]{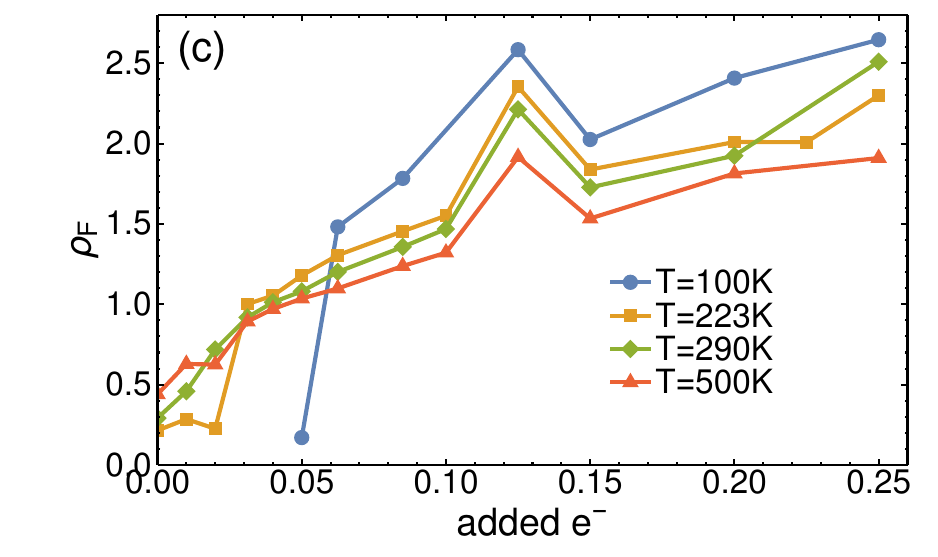}%
\includegraphics[width=0.5\textwidth]{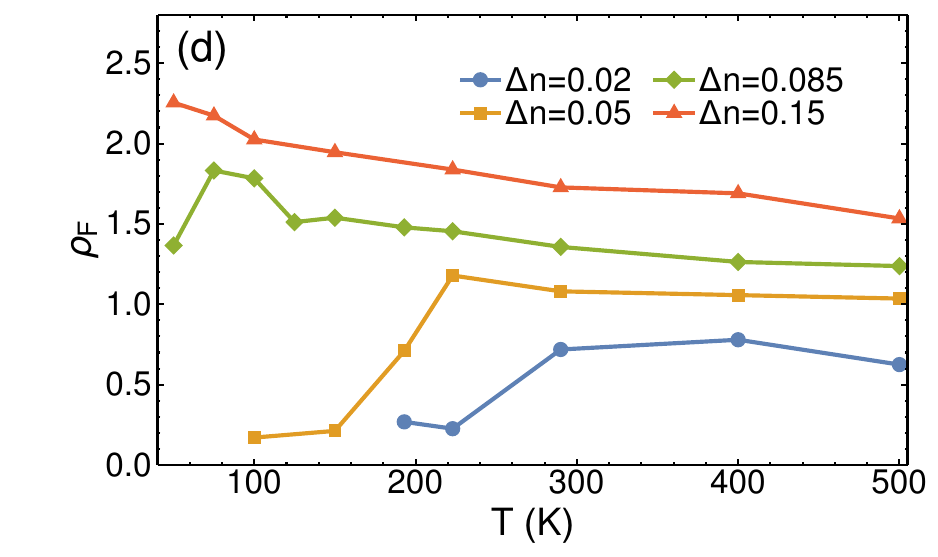}
\caption[Z and $\rho_F$ along constant $\Delta n$/temperature cuts ]{The quasi-particle spectral weight $Z$ (top) and the proxy for the DOS at the Fermi level $\rho_F$ (bottom) along cuts of constant temperature $T$ (left) and constant doping $\Delta n$ (right) in the $\Delta n$-temperature phase diagram of Fig. \ref{fig:phaseDiagram}.}
\label{fig:cuts}
\end{figure*}

Fig.~\ref{fig:cuts} shows selected line cuts through the phase diagrams in Fig.~\ref{fig:phaseDiagram}. While the fixed-temperature $Z$ curves depend sensitively on the doping level (Fig.~\ref{fig:cuts}a), the DOS ($\rho_F$) curves are relatively insensitive to temperature, particularly at higher temperatures  in the metallic phase (Fig.~\ref{fig:cuts}c). However, both the temperature at which $Z$ drops to zero (Fig.~\ref{fig:cuts}b) and the temperature at which there is a sudden drop in $\rho_F$ (Fig.~\ref{fig:cuts}d) clearly depend on the value of $\Delta n$, with no such drops being observed in either quantity above $\Delta n \sim 0.06$. 

\section{Conclusions}
Recent developments in electrostatic gating opened new avenues of research where the carrier concentrations in materials can be tuned continuously without necessarily introducing disorder or stoichiometric effects. Here, we used first principles DFT+eDMFT to study an filling induced metal-insulator transition in the Mott insulating transition metal sulfide NiS$_2$. Although DFT+eDMFT provides a more realistic setting than DMFT alone, there are still phenomena, such as polarons\cite{Capone2004} and Wigner localization\cite{Radonjic2012}, that we are not able to capture. If these effect suppress Mottness, they could possibly lead to an even lower carrier concentration needed to induce the metallic state. 
Our calculations show that the spectral properties behaves very differently when a Mott transition is induced by changing carrier concentration (filling control), rather than isovalent chemical substitution (bandwidth control). Increasing the carrier concentration leaves the system with significant incoherent spectral weight, while bandwidth manipulation via Se substitution creates coherent bands and minimal incoherent weight at the Fermi level. Our results further show that metal-insulator transitions at a wide range of temperatures can be induced in NiS$_2$ under modest doped electron concentrations. A concentration of $\Delta n = 0.06$ additional electrons per Ni atom corresponds to a surface electron density of $\sim\! 8 \! \times\! 10^{13}~ e^-/\mathrm{cm}^{2}$, assuming the gating penetrates exactly one unit cell, well within the capabilities of electrostatic gating \cite{Leighton2019}. Our results show that electrostatic gating can provide a complementary means to study the Mott transition via introducing carriers with minimum disorder. 

\acknowledgements
We thank C. Leighton and J. Walter for fruitful discussions. This work was supported primarily by the National Science Foundation through the University of Minnesota MRSEC under Award Number DMR-2011401.

\end{document}